\documentclass[aps,prb,twocolumn,showpacs]{revtex4}
\usepackage{bm}
\usepackage{amsmath}
\usepackage{amssymb}
\usepackage{graphicx}

\newcommand{\beq}{\begin{equation}}
\newcommand{\eeq}{\end{equation}}
\newcommand{\beqar}{\begin{eqnarray*}}
\newcommand{\eeqar}{\end{eqnarray*}}
\newcommand{\ETh}{E_{\text{Th}}}

\newcommand{\sgn}{\text{sgn}\,}

\newcommand{\Db}{\bar{D}}

\newcommand{\sea}{\searrow}
\newcommand{\nea}{\nearrow}
\newcommand{\swa}{\swarrow}
\newcommand{\nwa}{\nwarrow}

\newcommand{\Hh}{\hat{H}}

\newcommand{\Cc}{\mathcal{C}}
\newcommand{\Gc}{\mathcal{G}}

\newcommand{\Vc}{\mathcal{V}}

\newcommand{\Hb}{{\bf H}}
\newcommand{\nb}{{\bf n}}
\newcommand{\tb}{{\bf t}}

\newcommand{\ib}{{\bf i}}
\newcommand{\jb}{{\bf j}}
\newcommand{\ab}{{\bf a}}
\newcommand{\bb}{{\bf b}}
\newcommand{\eb}{{\bf e}}

\newcommand{\s}{{\bf s}}

\newcommand{\xb}{{\bf x}}
\newcommand{\qb}{{\bf q}}
\newcommand{\pb}{{\bf p}}

\newcommand{\Ab}{{\bf A}}
\newcommand{\rb}{{\bf r}}
\newcommand{\sbf}{{\bf s}}

\newcommand{\dg}{\dagger}

\newcommand{\lan}{\langle}
\newcommand{\ran}{\rangle}
\newcommand{\om}{\omega}

\newcommand{\al}{\alpha}
\newcommand{\be}{\beta}
\newcommand{\ga}{\gamma}
\newcommand{\Ga}{\Gamma}

\newcommand{\de}{\delta}

\newcommand{\sig}{\sigma}

\newcommand{\eps}{\varepsilon}

\newcommand{\lt}{\left}
\newcommand{\rt}{\right}
\begin{document}

\title{Hall Effect in Granular Metals: Weak Localization Corrections}
\author{Maxim Yu. Kharitonov$^{1}$ and Konstantin B. Efetov$^{1,2}$}
\affiliation{$^{1}$ Theoretische Physik III, Ruhr-Universit\"{a}t Bochum, Germany,\\
$^{2}$L.D. Landau Institute for Theoretical Physics, Moscow, Russia.}
\date{\today}

\begin{abstract}

We study the effects of localization on the Hall transport in
a granular system at large tunneling conductance $g_{T}\gg 1$
corresponding to the metallic regime.
We show that the first-order in $1/g_T$ weak localization correction to Hall resistivity
of a two- or three-dimensional granular array vanishes identically, $\de \rho_{xy}^{WL}=0$.
This result is in agreement with the one
for ordinary disordered metals.
Being due to an exact cancellation, our result holds
for arbitrary relevant values of temperature $T$ and magnetic field $H$,
both in the ``homogeneous'' regime of very low $T$ and $H$  corresponding
to  ordinary disordered metals
and in the ``structure-dependent'' regime of higher values of $T$ or $H$.

\end{abstract}
\maketitle

\section{Introduction}

Dense-packed arrays of
metallic or semiconducting nanoparticles imbedded into
an insulating matrix, usually called {\em granular systems}
or {\em nanocrystals},
form a new important class of artificial materials
with tunable electronic properties.
A lot about the transport and thermodynamical properties of such systems
has been already understood theoretically
(see a review \onlinecite{BELVreview} and references therein).
%
For instance, the longitudinal conductivity (resistivity) has
been calculated in both the metallic and insulating regimes.

At the same time, little attention has been paid to the Hall
transport in  such granular systems. Measuring the Hall
resistivity one can obtain important information about the system
and such a study is certainly desirable for the characterization
of  granular materials.

Hall transport in granular materials has been addressed
theoretically only recently in Refs.~\onlinecite{KEshort,KElong}.
In these works we studied the Hall effect in a granular system in
the metallic regime ({\em``granular metal''}), when the intergrain
tunneling conductance $G_T=(2e^2/\hbar) g_T$ is large, $g_T \gg 1$
(further we set $\hbar=1$). We have shown that at high enough
temperatures the Hall resistivity of a granular metal is given by
an essentially classical Drude-type expression \beq
    \rho^{(0)}_{xy}=\frac{H}{n^*ec},
\label{eq:rhoxy00}
\eeq
where the effective carrier density $n^*$ of the
system differs from the actual carrier density $n$
in the grains only by a numerical factor dependent
on the grain geometry and type of the granular lattice.
For a granular film, its sheet Hall resistance is obtained by
dividing Eq.~(\ref{eq:rhoxy00}) by the film thickness.

As the temperature $T$ is lowered, effects of Coulomb interaction
become especially important and can influence the transport
properties of the system significantly. Indeed, we have
demonstrated\cite{KEshort,KElong} that in quite a broad range  of
temperatures the classical Hall resistivity (\ref{eq:rhoxy00}) of
both two- (2D) and three- (3D) dimensional granular arrays
acquires a noticeable logarithmic correction due to the Coulomb
interaction, which is of local origin and absent in ordinary
homogeneously disordered metals.

The Coulomb interaction, however, is not the only source of
quantum contributions. Another quantum effect setting in at
sufficiently low temperatures is  weak localization (WL), which is
due to the interference of electrons moving along
self-intersecting trajectories. The first order in the inverse
tunneling conductance $1/g_T$ WL correction $\de \rho_{xx}^{WL}$
to the longitudinal resistivity of a granular metal, including its
dependence on the magnetic field $H$
(magnetoresistance)\cite{BCTV,BVG}, was studied in
Refs.~\onlinecite{BelUD,BCTV,BVG}. Being divergent for
two-dimensional samples\footnote{ Considering  Hall transport we
do not discuss the one-dimensional case of granular ``wires'' in
this paper.} (granular films consisting of one of a few grain
monolayers), the WL correction $\de \rho_{xx}^{WL}$ exhibits a
universal behavior at lowest temperatures $T$ and magnetic fields
$H$, in agreement with the theory of ordinary homogeneously
disordered metals. As $T$ or $H$ are increased or if the sample is
three-dimensional, the correction $\de \rho_{xx}^{WL}$ becomes
dependent on the granular structure of the system. In the latter
regime, however, the relative correction is already quite small
and does not exceed $1/g_T$.

In this work we study the effects of weak localization
on the Hall transport in  a granular system in the metallic regime.
We calculate first-order in $1/g_T$ weak localization corrections
to the Hall conductivity and resistivity
and find  that both for  2D and 3D arrays the
correction  to the Hall resistivity vanishes identically:
\[
    \de\rho_{xy}^{WL} = 0.
\]
This result is in agreement with the one obtained for
homogeneously disordered metals in Ref.~\onlinecite{Fukuyama,AKLL}.
Being due to an exact cancellation, it holds for arbitrary values
of temperature and magnetic field, both in the ``homogeneous''
regime of very low $T$ and $H$ and in the ``structure-dependent''
regime of higher values of $T$ or $H$.
Of course, this cancellation occurs under certain assumptions, but
they are the same as those under
which a nonvanishing correction
$\de\rho_{xx}^{WL}$ to the longitudinal resistivity was obtained\cite{BelUD,BCTV,BVG}.


\section{Model and method}

The model we use in this paper is essentially the same as the one
studied in Refs.~\onlinecite{KEshort,KElong}
and we refer the reader to those works for details.
We consider a quadratic (2D, $d=2$) or cubic (3D, $d=3$) lattice
of metallic grains coupled to each other by tunnel contacts (see
Fig.~\ref{fig:main}) and assume translational symmetry of the
lattice, i.e., equal conductances $G_T$ of all contacts and
identical properties of all grains (size and shape, density of
states, etc.). To simplify the calculations further, we also
assume the intragrain electron dynamics diffusive, i.e., that the
mean free path $l$ is much smaller than the size $a$ of the grain
($l \ll a$). However, our results are also valid for ballistic ($l
\gtrsim a$) intragrain disorder provided the electron intragrain
motion is classically chaotic. In the metallic regime ($g_T \gg
1$) the localization effects can be studied perturbatively in
$1/g_T$, as long as the relative corrections remain small. We
perform calculations for magnetic fields $H$, such that $\om_H
\tau_0 \ll 1$, where $\om_H = e H/(mc)$ is the cyclotron frequency
and $\tau_0$ is the  electron scattering time inside the grain.
The condition $\om_H \tau_0 \ll 1$ is well met for granular arrays
even for experimentally high fields owing to the small size $a$ of
the grains. We assume that the granularity of the system is
``well-pronounced'', i.e., that the condition \beq
    \Ga \ll \ETh,
\label{eq:granular}
\eeq
is fulfilled, where $\Ga$ is the tunneling escape rate and $\ETh$ is the Thouless energy of the grain.


We write the Hamiltonian describing the system as
\beq
    \Hh = \Hh_0 + \Hh_t + \Hh_c.
\label{eq:H}
\eeq

In Eq.~(\ref{eq:H}),
\beq
    \Hh_0= \sum_\ib \int d\rb_\ib \psi^\dg(\rb_\ib)
    \lt\{ \xi\lt[\pb_\ib-\frac{e}{c} \Ab(\rb_\ib)\rt]+ U(\rb_\ib) \rt\} \psi(\rb_\ib)
\label{eq:H0}
\eeq
is the electron Hamiltonian of isolated grains,
$\xi(\pb)=\pb^2/2m-\epsilon_F$,
$\Ab(\rb_\ib)$ is the vector potential describing
the uniform magnetic field $\Hb=H\eb_z$ pointing in the $z$ direction,
$U(\rb_\ib)$ is the random disorder potential of the grains,
$\ib=(i_1, \ldots , i_d)$ is an integer vector numerating the grains.
The disorder average is performed using a Gaussian distribution with the variance
\beq
    \lan U(\rb_\ib) U(\rb'_\ib) \ran_U = \frac{1}{2 \pi \nu \tau_0} \de(\rb_\ib-\rb'_\ib),
\label{eq:UU}
\eeq
where $\nu$ is the density of states in the grain at the Fermi level per one spin projection
and $\tau_0$ is the intragrain scattering time.

Further, the tunneling Hamiltonian in Eq.~(\ref{eq:H}) is given by
\beq
    \Hh_t= \sum_{\lan \ib,\jb \ran} (X_{\ib,\jb}+X_{\jb,\ib}), \mbox{ }
        X_{\ib,\jb}=\int d\s_\ib d\s_\jb \,
        t(\s_\ib, \s_\jb) \psi^\dg(\s_\ib) \psi(\s_\jb),
\label{eq:Ht}
\eeq
where the summation is taken over the neighboring grains connected by a
tunnel contact, and the integration is done over the surfaces of
the contact.
The tunneling amplitudes $t(\s_\ib, \s_\jb)
$
are assumed to be random Gaussian variables with the variance
\beq
    \lan t(\s_\ib, \s_\jb) t(\s_\jb, \s_\ib) \ran_t = t_0^2 \de(\s_\ib-\s_\jb),
\label{eq:tt}
\eeq
where $\de(\sbf_\ib-\sbf_\jb)$ is a $\de$-function on the contact surface,
and $t_0^2$ has a meaning of tunneling probability
per unit area of the contact. This accounts for inevitable irregularities
of the tunnel barriers on atomic scales
and well models the local nature of tunneling between metallic grains.

Finally, the last term $\Hh_c$ in Eq.~(\ref{eq:H})
stands for the Coulomb interaction between the electrons.
In the leading first in $1/g_T$ order the Coulomb
interaction results in the phase relaxation
yielding a finite dephasing rate $1/\tau_\phi$ in the Cooperon self-energy\cite{BVG,BelUD}.
Since our main result does not depend on the explicit form of the Cooperon,
we will not deal with the Coulomb interaction in this paper
and will omit $\Hh_c$ in Eq.~(\ref{eq:H}) from now on.

The conductivity is calculated using the Kubo formula
in Matsubara technique\cite{AGD}:
\beq
    \sig_{\ab\bb}(\om)= 2e^2 a^{2-d}\frac{1}{\om}
        \sum_\jb \lt[\Pi_{\ab\bb}(\om, \ib-\jb)-\Pi_{\ab\bb}(0, \ib-\jb)\rt]
\label{eq:Sig}
\eeq
where $\om \in 2\pi T {\mathbb Z} $  is a bosonic Matsubara frequency
(${\mathbb Z}$ is a set of integers, throughout the paper we assume $\om \geq 0$),
$\ab$ and $\bb$ are the lattice unit vectors, and
\[
    \Pi_{\ab\bb}(\om,\ib-\jb)=\int_0^{1/T} d\tau\,e^{i \om \tau}
        \Pi_{\ab\bb}(\tau,\ib-\jb),
\]
\beq
    \Pi_{\ab\bb}(\tau,\ib-\jb)=\lan T_\tau I_{\ib,\ab}(\tau) I_{\jb,\bb}(0) \ran
\label{eq:Pi}
\eeq
is the current-current correlation function.
In Eq.~(\ref{eq:Pi}),
\beq
    I_{\ib,\ab}(\tau) =X_{\ib+\ab,\ib}(\tau)-X_{\ib,\ib+\ab}(\tau),
\label{eq:I}
\eeq
the thermodynamic average $\lan\ldots\ran$ is taken with
the Hamiltonian $\Hh=\Hh_0+\Hh_t$ [Eq.~(\ref{eq:H}) with discarded $\Hh_c$], and
$A(\tau)=e^{\Hh\tau}A e^{-\Hh \tau}$ is the Heisenberg representation of any operator $A$.

\section{Weak localization corrections}

\begin{figure*}
\includegraphics{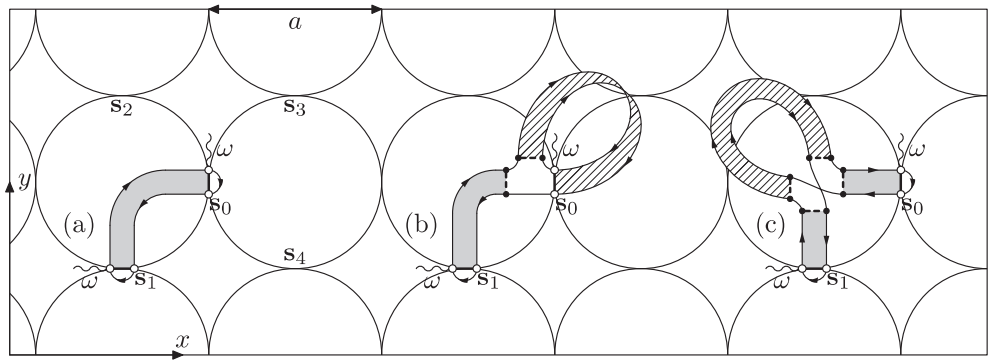}
\caption{
\label{fig:main}
(a) Diagrams for the ``bare'' classical Hall conductivity $\sig_{xy}^{(0)}$
[Eqs.~(\ref{eq:sigxy0}) and (\ref{eq:RH})] of the granular metal.
The contact $\s_1$ is connected to the contact $s_0$ by the intragrain diffuson (gray stripe).
Same contributions from the contacts $\s_2, \s_3, \s_4$
have to be taken into account [Eq.~(\ref{eq:RH})].
(b) Diagrams describing the weak localization correction
to the conductance $G_T$ of the tunnel contact.
A diagram with the Cooperon $\Cc_0$ (rendered with lines stripe) flipped down and
the same two diagrams for the second contact also have to be considered.
(c) An example of a diagram for the weak localization correction to the
Hall resistance $R_H$ of the grain, which is expressed through the intragrain
diffuson $D(\om,\rb,\rb')$ [Eqs.~(\ref{eq:D}) and (\ref{eq:Dbc})].
The diagram contributes to the renormalization of the
diffusion coefficient $D_0$ [Eq.~(\ref{eq:D})], the complete set of such diagrams shown in Fig.~\ref{fig:dD0}.
Weak localization corrections to the boundary condition (\ref{eq:Dbc})
must be also taken into account (see Fig.~\ref{fig:djr}).
}
\end{figure*}

The ``bare'' (i.e., without quantum effects)  Hall conductivity
of a granular metal is given by\cite{KEshort,KElong}
\beq
    \sig^{(0)}_{xy}= G_T^2 R_H a^{2-d},
\label{eq:sigxy0}
\eeq
where $G_T$ is the conductance of the tunnel contact and
$R_H$ is the Hall resistance of the grain. The latter is expressed through
the intragrain diffuson as
\beq
    R_H=\frac{1}{2e^2 \nu} (\Db_\nea -\Db_\sea + \Db_\swa - \Db_\nwa),
\label{eq:RH}
\eeq
where
\[
\bar{D}_\al=\frac{1}{S_0^2} \int d\s_0 d\s_a \bar{D}(\s_0,\s_a),
\]
with $a=1,2,3,4$ for $\al=\nea,\sea,\swa,\nwa$, respectively [see Fig.~\ref{fig:main}(a)],
$S_0$ is the area of the contact,
and $\Db(\rb,\rb')$ is the diffusion propagator of a single grain at $\om=0$
with discarded zero mode $1/(\om \Vc)$ ($\Vc$ is the grain volume).
We specify $\Db(\rb,\rb')$ explicitly somewhat later.
The essentially classical result (\ref{eq:sigxy0}) is given
 by the diagram in Fig.~\ref{fig:main}(a),
in which the contacts $\s_a$, $a=1,2,3,4$,  are connected to the contact $\s_0$
by the intragrain diffuson $\Db(\rb,\rb')$.
In order not to overcomplicate the calculations we
consider the range of frequencies $\om \ll \ETh$ in this paper,
which allows us to neglect the intragrain Coulomb interaction
when calculating the bare Hall conductivity (see Ref.~\onlinecite{KElong} for details).

We emphasize the crucial for the Hall effect technical point\cite{KEshort,KElong}:
the nonvanishing contribution
to the Hall conductivity [Eqs.~(\ref{eq:sigxy0}) and (\ref{eq:RH})]
comes from {\em nonzero modes} of the diffuson $\Db(\rb,\rb')$ only, whereas
the zero mode $1/(\om \Vc)$ simply drops out
due to the sign structure of Eq.~(\ref{eq:RH}).

Since the bare longitudinal conductivity equals
\beq
    \sig_{xx}^{(0)}= a^{2-d} G_T,
\label{eq:sigxx0}
\eeq
the bare Hall resistivity, following from Eqs.~(\ref{eq:sigxy0}) and  (\ref{eq:sigxx0}),
\beq
    \rho^{(0)}_{xy}= \frac{\sig^{(0)}_{xy}}{\lt(\sig^{(0)}_{xx}\rt)^2} = R_H a^{d-2}
\label{eq:rhoxy0}
\eeq
is independent of the intergrain tunneling conductance $G_T$.
It can be further shown\cite{KEshort,KElong}
that the Hall resistance $R_H$ [Eq.~(\ref{eq:RH})] is
{\em independent} of the scattering time $\tau_0$
and Eq.~(\ref{eq:rhoxy0}) leads to
Eq.~(\ref{eq:rhoxy00}).

In the first order in the inverse tunneling conductance $1/g_T$,
the weak localization corrections to the classical result (\ref{eq:sigxy0})
are given by the sum of all ``minimally crossed'' diagrams.
The ``fan-shaped'' ladder arising in such diagrams
corresponds to the well-known particle-particle propagator called {\em ``Cooperon''},
which can be formally defined for a granular metal in the same
way as for an ordinary disordered metal:
\beq
    \Cc(\om,\rb_\ib,\rb_\jb')=\frac{1}{2 \pi \nu}
        \lan \Gc(\eps+\om,\rb_\ib,\rb_\jb')
        \   \Gc(\eps,\rb_\ib,\rb_\jb') \ran_{U,t},
            \mbox{ } (\eps+\om)\eps < 0.
\label{eq:Cc}
\eeq
Here $\Gc$'s are the ``exact'' Green functions in the Matsubara technique
and the average is taken over the intragrain and tunnel contact disorder
with the help of Eqs.~(\ref{eq:UU}) and (\ref{eq:tt}).
The points $\rb_\ib$ and $\rb_\jb'$ may belong to
arbitrary distant grains $\ib$ and $\jb$.

One can calculate
the Cooperon $\Cc(\om,\rb_\ib,\rb_\jb')$
using the same diagrammatic rules as those for the diffuson\cite{KElong}.
They are governed by the condition $p_F a \gg 1$
($p_F$ is the Fermi momentum in the grains)
that each grain is a ``good'' metallic sample.
This demands that
the diagrammatic ``paths'' of the Green functions
$\Gc(\eps+\om,\rb_\ib,\rb_\jb')$ and  $\Gc(\eps,\rb_\ib,\rb_\jb')$
through intermediate grains and  contacts {\em coincide}.
Therefore,
the full Cooperon (\ref{eq:Cc}) is ``composed''
of the Cooperons
\beq
    C(\om,\rb,\rb')=\frac{1}{2 \pi \nu}
        \lan \Gc(\eps+\om,\rb,\rb')
        \   \Gc(\eps,\rb,\rb') \ran_{U},
            \mbox{ } (\eps+\om)\eps < 0
\label{eq:C}
\eeq
of isolated grains.
In Eq.~(\ref{eq:C}), $\rb$ and $\rb'$ belong to the same given grain
and tunneling to the neighboring grains should be completely neglected.

Although in order to obtain  nonvanishing Hall conductivity (\ref{eq:sigxy0}),
one is forced to take nonzero modes
in the intragrain diffuson $\Db(\rb,\rb')$ into account\cite{KEshort,KElong},
the zero modes in the Cooperons themselves do not drop out
from the expressions for WL corrections.
Therefore due to the
small size of the grains
one may
use the ``zero-mode'' approximation
for the Cooperons, i.e., to leave only the zero mode $1/(\om \Vc)$ in each grain in the expression
for the Cooperon (\ref{eq:C}).
To do so, however, the condition (\ref{eq:granular}) alone is not sufficient,
since the Cooperons are sensitive to magnetic field,
and in the presence of magnetic field
an additional condition must be met. Namely, the magnetic flux $H a^2$
 threading through
each grain must be smaller than the flux quantum $c/e$:
\beq
    \frac{e}{c}H a^2 \ll 1.
\label{eq:smallflux}
\eeq

Under the conditions (\ref{eq:granular}) and (\ref{eq:smallflux})
the spatial dependence of the intragrain Cooperon (\ref{eq:C})
coming from nonzero modes can be neglected and one gets:
\[
    C(\om,\rb,\rb') \approx \frac{1}{\Vc} \frac{1}{\om+{\cal E}(H)},
\]
where ${\cal E}(H) \propto D_0 (\frac{e}{c} H a)^2$ is the ``mass term''
acquired due to dephasing  by the magnetic field within the grain
[$D_0$ is the intragrain diffusion coefficient defined after Eq.~(\ref{eq:D})].
After that, the Cooperon
$\Cc(\om,\rb_\ib,\rb_\jb')$ [Eq.~(\ref{eq:Cc})]
of the whole granular system
depends on the grain indices $\ib$ and $\jb$ only and
we denote such ``zero-mode'' Cooperon as $\Cc_0(\om,\ib,\jb)$.
Its properties in the presence of magnetic field were studied in
Refs.~\onlinecite{BCTV,BVG}. Since
our main result, the vanishing WL correction to the Hall resistivity,
does not depend on the explicit form of $\Cc_0(\om,\ib,\jb)$,
we do not repeat
these properties
here,
reminding for reference only
that in the absence of magnetic field and dephasing effects one has
\[
   \Cc_0(\om,\ib,\jb)=\int \frac{a^d d^d \qb}{(2\pi)^d}
\frac{ e^{i a \qb (\ib-\jb)}  }{\om+2 \Ga \sum_\be [1-\cos(q_\be a)]},
\]
where the integration is done over the first Brillouin zone $\qb \in [-\pi/a,\pi/a]^d$
and $\be=x,y$ in 2D and $\be=x,y,z$ in 3D.
Note, that we have removed the inverse grain volume $1/\Vc$
from  the definition of $\Cc_0(\om,\ib,\jb)$.

We can now proceed with calculations of the weak localization corrections.
Conveniently, the contributions from the diagrams giving
first-order 
corrections
to HC $\sig_{xy}^{(0)}$ are factorized
according to the structure of Eq.~(\ref{eq:sigxy0}),
i.e., each diagram can be attributed to the renormalization of either
the tunneling conductance $G_T$ of the contact or the Hall resistance $R_H$
of the grain.
Below we study these two types of corrections separately.

\subsection{Weak localization correction to $G_T$}
First consider the diagram in Fig.~\ref{fig:main}(b).
In this diagram the Cooperon $\Cc_0(\om,\ib+\eb_x,\ib)$
connects the points belonging to two sides (in the grains $\ib+\eb_x$ and $\ib$)
of the same contact $(\ib+\eb_x,\ib)$.
Note that such diagrams arise form the ``particle-particle pairing''
$X_{\ib+\eb_x,\ib}(\tau) X_{\ib+\eb_x,\ib}(\tau_1)$ [see Eqs.~(\ref{eq:Ht}) and (\ref{eq:I})]
of the tunneling operators at the considered contact $(\ib+\eb_x, \ib)$,
whereas in the diagram in Fig.~\ref{fig:main}(a) for the bare
conductivity  we have ``particle-hole pairing''
$X_{\ib+\eb_x,\ib}(\tau) X_{\ib,\ib+\eb_x}(\tau_1)$.

Since the other elements of the diagram in Fig.~\ref{fig:main}(b)
remain unaffected, this diagram can be attributed
to the renormalization of the conductance $G_T$ of the tunnel contact
in Eq.~(\ref{eq:sigxy0}). Indeed, considering the same diagrams
for the other contact in Fig.~\ref{fig:main}(a),
for the relative correction to HC $\sig_{xy}^{(0)}$ [Eq.~(\ref{eq:sigxy0})]
we obtain:
\beq
    \frac{  \de \sig_{xy}^{(1)}(\om)}{ \sig_{xy}^{(0)}}= 2 \frac{\de G_T(\om)}{G_T}
\label{eq:dsigxy1},
\eeq
where
\beq
    \frac{\de G_T(\om)}{G_T} =\frac{1}{2 \pi \nu \Vc}
    [\Cc_0(\om,\ib+\ab,\ib)+\Cc_0(\om,\ib,\ib+\ab)],
\label{eq:dGT} \eeq and $\ab=\eb_x$ or $\ab=\eb_y$, [assuming the
square/cubic symmetry of the lattice, we do not distinguish
between $x$ and $y$ directions]. In Eq.~(\ref{eq:dsigxy1}), the
factor $2$ stands for two contacts according to the square $G_T^2$
in Eq.~(\ref{eq:sigxy0}). As expected, the expression
(\ref{eq:dGT}) for the relative correction to $G_T$ obtained from
the diagrams in Fig.~\ref{fig:main}(b) coincides with the one
obtained from calculating WL correction to the longitudinal
conductivity $\sig_{xx}^{(0)}$ in
Refs.~\onlinecite{BelUD,BCTV,BVG}: \beq
    \frac{  \de \sig_{xx}^{WL}(\om)}{ \sig_{xx}^{(0)}}= \frac{\de G_T(\om)}{G_T}
\label{eq:dsigxx}.
\eeq

Since the correction (\ref{eq:dsigxy1}) contributes solely to the renormalization of
the tunneling conductance $G_T$ and the bare HR $\rho_{xy}^{(0)}$ [Eq.~(\ref{eq:rhoxy0})]
simply does not contain $G_T$, the corresponding WL correction to HR from
the diagrams in Fig.~\ref{fig:main}(b) vanishes:
\beq
    \frac{  \de \rho_{xy}^{(1)}(\om)}{ \rho_{xy}^{(0)}}=
    \frac{  \de \sig_{xy}^{(1)}(\om)}{ \sig_{xy}^{(0)}}-
    2\frac{ \de \sig_{xx}^{WL}(\om)}{ \sig_{xx}^{(0)}} \equiv 0.
\label{eq:drhoxy1}
\eeq

\subsection{Weak localization correction to $R_H$}
Now let us consider the diagram shown in Fig.~\ref{fig:main}(c).
This diagram describes the effect of localization on
the intragrain diffuson $D(\om,\rb,\rb')$ and, eventually,
contributes to the renormalization of the Hall resistance $R_H$ of the grain, expressed
through the diffuson according to Eq.~(\ref{eq:RH}).
The aim of this section is to show that the
WL correction to the Hall resistance (\ref{eq:RH})
arising from all such diagrams actually vanishes:
\beq
    \de R_H^{WL} =0.
\label{eq:dRH}
\eeq

We remind the reader that
the intragrain diffuson is defined formally as
\beq
    D(\om,\rb,\rb') \equiv \frac{1}{2 \pi \nu} \lan \Gc(\eps+\om,\rb,\rb')
            \Gc(\eps,\rb',\rb)\ran_U,
            \mbox{ } (\eps+\om)\eps < 0.
\label{eq:Ddef}
\eeq
where $\lan \ldots \ran_U$ denotes the averaging over the intragrain disorder
according to Eq.~(\ref{eq:UU}).
\subsubsection{Intragrain diffuson in the absence of  weak localization effects}
In the absence of weak localization effects
(i.e., in the ``noncrossing approximation''\cite{AGD})
the average (\ref{eq:Ddef}) is given by a series of ladder-type diagrams.
The summation of this series is equivalent to solving a certain
integral equation, which in the diffusive limit ($\om \tau_0 \ll 1$ and $l \ll a$)
can be reduced to a differential diffusion equation
\beq
    (\om- D_0 \nabla_\rb^2) D(\om,\rb,\rb')=\de(\rb-\rb').
\label{eq:D}
\eeq
Here $D_0=v_F l /3$ is the classical diffusion coefficient in the grain
($v_F$ is the Fermi velocity, $l=v_F \tau_0$ is the electron  mean free path,
$D_0$ is not affected by magnetic field, such that $\om_H\tau_0 \ll 1$).

\begin{figure}
\includegraphics{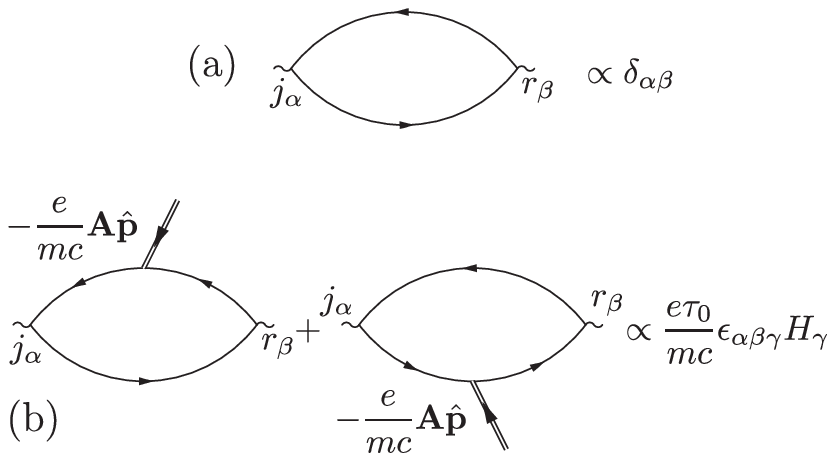}
\caption{\label{fig:jr0}
Diagrams for the current-coordinate correlation function $\lan j_\al r_\be \ran_0$
[Eqs.~(\ref{eq:jr0}) and (\ref{eq:jr0expr})] in the absence of weak localization.
Fermionic lines denote the Green function
$[G(\eps,\pb)]^{-1}=i \eps - \xi_\pb + \frac{i}{2 \tau_0} \sgn \eps$ of a bulk metal with $H=0$.
(a) Magnetic-field-independent part of  $\lan j_\al r_\be \ran_0$
giving the LHS of the boundary condition (\ref{eq:Dbc}).
(b) Linear in magnetic field part of $\lan j_\al r_\be \ran_0$
obtained by inserting the ``magnetic vertex'' $-\frac{e}{mc} \Ab \hat{\pb}$
in all possible ways into the diagram (a) and giving the RHS of Eq.~(\ref{eq:Dbc}).
}
\end{figure}

For a finite system (a grain), Eq.~(\ref{eq:D})
must be supplied by a proper boundary
condition.
In Ref.~\onlinecite{KElong}
we have shown that  the boundary condition
in the diffusive case has the form
\beq
    n_\al \lan j_\al r_\be \ran_0 \nabla_{\rb\be} D(\om,\rb,\rb') |_{\rb\in S}=0.
\label{eq:bcD}
\eeq
Here, the coordinate $\rb$ belongs to the grain boundary $S$,
the unit vector $\nb$ normal to the grain boundary points outside the grain,
$\al,\be=x,y,z$, and

\beq
    \lan j_\al r_\be \ran_0= \int d\xb \hat{j}_{\rb\al}
    [G(\eps+\om,\rb,\xb) G(\eps,\xb,\rb)]
        (\xb-\rb)_\be
\label{eq:jr0}
\eeq
is the current-coordinate correlation function.
In Eq.~(\ref{eq:jr0}), $G(\eps+\om,\rb,\xb) = \lan \Gc(\eps+\om,\rb,\xb) \ran_U$
is the disorder-averaged Green function of the grain and
the current operator $\jb_\rb$ acts on the product
of two Green functions as
\begin{eqnarray}
    \hat{\jb}_\rb [ G(\eps+\om,\rb,\xb) G(\eps,\xb,\rb)]  = \nonumber\\
    =\frac{1}{2m} [ G(\eps,\xb,\rb)
 (-i\nabla_\rb) G(\eps+\om,\rb,\xb) \nonumber\\
%
%
    + G(\eps+\om,\rb,\xb) (i\nabla_\rb) G(\eps,\xb,\rb)] \nonumber \\
    - \frac{e}{mc} \Ab(\rb) G(\eps+\om,\rb,\xb)
 \Gc(\eps,\xb,\rb),
\label{eq:joper}
\end{eqnarray}
where the vector potential $\Ab(\rb)$ corresponds to the magnetic field $H$.
%


Owing to the small spatial scale ($\sim l\ll a$)
of the kernel in Eq.~(\ref{eq:jr0}),
$\lan j_\al r_\be \ran_0$ may be evaluated
for $\rb$ located not directly on the grain boundary,
but a few $l$ away from it
in the bulk of the grain, where
the expressions for the bulk can be used for $G$'s.
Note that the ladder contribution to $\lan j_\al r_\be \ran_0$
vanishes in the case of the white noise-disorder [Eq.~(\ref{eq:UU})].

As we are study Hall transport, the correlation  function $\lan j_\al r_\be \ran_0$
has to be calculated taking the magnetic field $H$ into account,
which may be done in the linear in $H$ order,
since the condition $\om_H \tau_0 \ll 1$ is assumed to be met.
The calculations 
can be performed
with the help of the  diagrammatic technique either by directly
expanding Green functions $G$ in the vector potential $\Ab(\rb)$ or
using an explicitly gauge-invariant approach developed by Khodas and Finkel'stein
in Ref.~\onlinecite{Khodas}.
We choose the former approach here.
The diagrams for $\lan j_\al r_\be \ran_0$ in the absence of WL effects
are given in Fig.~\ref{fig:jr0} and we obtain\cite{KElong}:
\beq
    \lan j_\al r_\be \ran_0 = \Lambda \lt(
    \de_{\al\be} + \frac{e\tau_0}{m c} \epsilon_{\al\be\ga} H_\ga \rt),
\label{eq:jr0expr}
\eeq
where $\epsilon_{\al\be\ga}$ is the totally antisymmetric tensor, $\epsilon_{xyz}=1$,
and $\Lambda=-(2\pi/3)\nu l^2$
is an irrelevant for the boundary condition (\ref{eq:bcD}) prefactor.
Inserting Eq.~(\ref{eq:jr0expr}) into Eq.~(\ref{eq:bcD}), we get
\beq
    (\nb, \nabla_\rb D)|_{\rb \in S}= \om_H  \tau_0 (\tb, \nabla_\rb D)|_{\rb \in S},
\label{eq:Dbc}
\eeq
where $\tb = [\nb,\Hb]/H$ is the tangent vector
pointing in the direction opposite to the edge drift.

The propagator $\Db(\rb,\rb')$ entering the expression (\ref{eq:RH})
for the Hall resistance $R_H$ of the grain, satisfies Eqs.~(\ref{eq:D}) and (\ref{eq:Dbc})
with $\om=0$, i.e.,
$\Db(\rb,\rb')$ is a Green function for the Poisson equation
with the boundary condition (\ref{eq:Dbc}).

\subsubsection{Intragrain diffuson renormalized by  weak localization effects}

We can now proceed with WL corrections to the intragrain diffuson $D(\om,\rb,\rb')$.
Since neglecting localization effects
the diffuson $D(\om,\rb,\rb')$ [Eq.~(\ref{eq:Ddef})]
has been reduced to the solution
of Eqs.~(\ref{eq:D}) and (\ref{eq:bcD}), our task now
is to find out how these equations are affected by weak localization.
It is very important that for a finite system with boundary (grain)
one has to renormalize not only the diffusion equation (\ref{eq:D}) itself,
but also the boundary condition (\ref{eq:bcD}) for the diffuson.

\begin{figure*}
\includegraphics{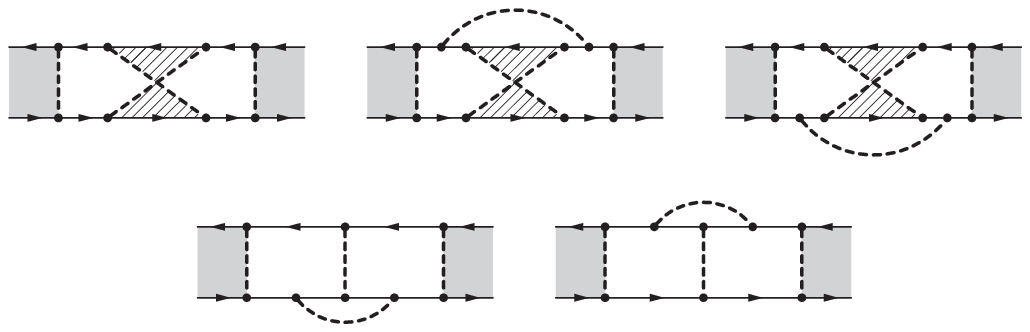}
\caption{\label{fig:dD0}
Diagrams for the weak localization correction to the diffusion coefficient
$D_0$ [Eqs.~(\ref{eq:D0tilde}) and (\ref{eq:c})]
of the intragrain diffuson $D(\om,\rb,\rb')$ (gray blocks).
Diagrams in the upper row form a Hikami box,
the twisted rendered with lines block denotes the Cooperon $\Cc_0(\om,\ib,\ib)$.
Diagrams in the lower row are of the same order as the sum of those
in the upper row and are missing in the ladder summation for $\Cc_0(\om,\ib,\ib)$.
}
\end{figure*}

\begin{figure*}
\includegraphics{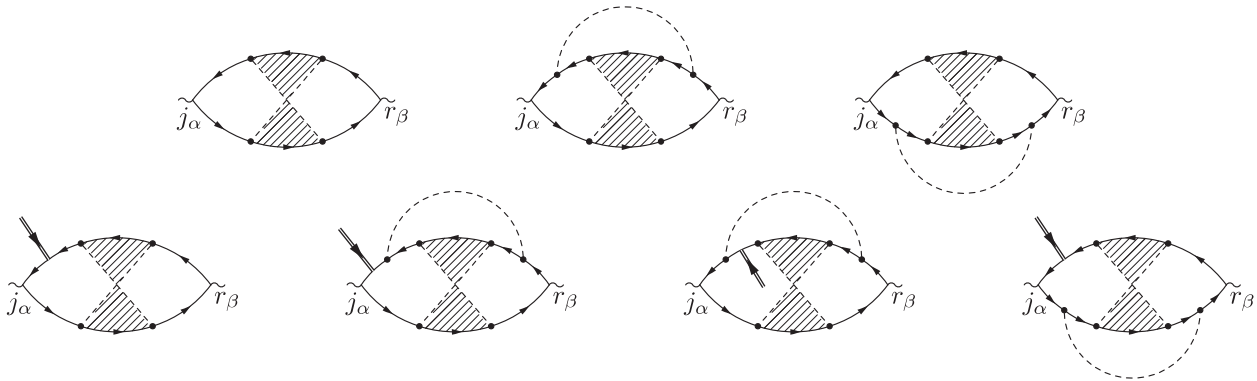}
\caption{\label{fig:djr}
Diagrams for the weak localization corrections [Eq.~(\ref{eq:jr})]
to the current-coordinate correlation function $\lan j_\al r_\be \ran_0$
[Eqs.~(\ref{eq:jr0}) and (\ref{eq:jr0expr})].
Diagrams in the upper row forming a Hikami box
describe the correction to the magnetic-field-independent
part of $\lan j_\al r_\be \ran_0$ [Fig.~\ref{fig:jr0}(a)].
Diagrams in the lower row describe the correction
to the linear in the magnetic field part of $\lan j_\al r_\be \ran_0$ [Fig.~\ref{fig:jr0}(b)].
Similar insertions of the magnetic vertex into the right block of Green functions
and the upside-down flip of such diagrams must also be considered.
}
\end{figure*}

We start by considering  the diffusion equation~(\ref{eq:D}).
In a bulk metal effects of localization on the diffusive electron motion
were first studied in Ref.~\onlinecite{GLK} by Gorkov, Larkin, and Khmelnitski.
It was shown that the diffusion equation (\ref{eq:D}) remains valid,
but the diffusion constant $D_0$ is renormalized.
The diagrams describing renormalization of $D_0$
are obtained by inserting the ``fan-shaped'' ladder
into the ordinary ladder describing the diffuson $D(\om,\rb,\rb')$,
as shown in Fig.~\ref{fig:dD0}.
Their calculation is more challenging for a granular system
due to the possibility of tunneling between the grains.
Nevertheless,
under the assumed conditions (\ref{eq:granular}) and (\ref{eq:smallflux})
we obtain a result essentially the same as that of Ref.~\onlinecite{GLK}
for the renormalized diffusion coefficient:
\beq
    \tilde{D}_0(\om)= D_0 [1-c(\om)],
\label{eq:D0tilde}
\eeq
where
\beq
    c(\om)= \frac{1}{\pi \nu \Vc} \Cc_0(\om,\ib,\ib)
\label{eq:c}
\eeq
is given by the zero-mode Cooperon with coinciding points.
Since the characteristic scale of the Cooperon is $\Cc_0(\om,\ib,\ib)
 \sim 1/\Ga$
and the mean level spacing in each grain is $\de = 1/(\nu \Vc)$,
the relative correction $c(\om) \sim \de/\Ga=1/g_T$
is proportional to the inverse intergrain conductance $1/g_T$.

More interestingly, for a finite system one also has to consider
the effect of WL on the boundary condition (\ref{eq:bcD}). The
sensitivity of the boundary condition to WL effects is crucial
for the Hall transport, since, as it was discussed in
Ref.~\onlinecite{KElong}, the differences $\Db_\nea -\Db_\sea$
and  $\Db_\swa - \Db_\nwa$ in Eq.~(\ref{eq:RH}) for $R_H$ are
nonvanishing solely
due to the presence of the magnetic field in 
Eq.~(\ref{eq:bcD}).
Since the boundary condition (\ref{eq:bcD}) is determined
by the correlation function (\ref{eq:jr0}),
we  need to find WL correction to this quantity.
The corresponding diagrams are shown in Fig.~\ref{fig:djr}.
Their calculation is somewhat cumbersome, but straightforward, and yields
the following result for the renormalized correlation function:
\beq
    \lan j_\al r_\be \ran= \Lambda \lt\{
    \de_{\al\be}[1-c(\om)] +
    \frac{e\tau_0}{m c} \epsilon_{\al\be\ga} H_\ga [1-2 c(\om)] \rt\},
\label{eq:jr}
\eeq
where $c(\om)$ is given by Eq.~(\ref{eq:c}).

As a result,
replacing $D_0$ by  $\tilde{D}_0(\om)$ [Eq.~(\ref{eq:D0tilde})]
in Eq.~(\ref{eq:D})
and  $\lan j_\al r_\be \ran_0$ by $\lan j_\al r_\be \ran$ [Eq.~(\ref{eq:jr})]
in Eq.~(\ref{eq:bcD}), we obtain that the renormalized diffuson satisfies
the equation
\beq
    \lt\{ \om- D_0[1-c(\om)] \nabla_\rb^2 \rt\}D(\om,\rb,\rb') = \de(\rb-\rb')
\label{eq:Drenorm}
\eeq
and the boundary condition
\beq
    (\nb, \nabla_\rb D)|_{\rb \in S}=
    \om_H \tau_0 [1-c(\om)] (\tb, \nabla_\rb D)|_{\rb \in S}.
\label{eq:bcDrenorm}
\eeq
instead of Eqs.~(\ref{eq:D}) and (\ref{eq:Dbc}), respectively.
In Eq.~(\ref{eq:bcDrenorm}) we put $[1-2c(\om)]/[1-c(\om)] \approx 1-c(\om)$,
since $c(\om) \ll 1$ within the validity of the perturbation approach.

\subsubsection{Vanishing weak localization correction to the grain
Hall resistance $R_H$}
Now let us see how the obtained renormalizations affect the Hall resistance
$R_H$ [Eq.~(\ref{eq:RH})] of the grain.
Although Eqs.~(\ref{eq:Drenorm}) and (\ref{eq:bcDrenorm}) cannot be solved
for an arbitrary shape of the grains, this is not actually necessary
and the needed conclusions about $R_H$ can be drawn based on
the following rather simple analysis.

The characteristic value of $\Db_\al$'s in Eq.~(\ref{eq:RH})
can be estimated from Eq.~(\ref{eq:Drenorm}) as
\beq
    \Db_\al \propto \frac{1}{a^3} \frac{1}{D_0 [1-c(\om)]/a^2}.
\label{eq:Dest}
\eeq
The differences $ \Db_\nea -\Db_\sea = \Db_\swa - \Db_\nwa$ in Eq.~(\ref{eq:RH}),
however, require a more accurate estimate, since they are nonzero
only in the presence of magnetic field $H \neq 0$
due to the directional asymmetry $\Db(\rb,\rb') \neq \Db(\rb',\rb)$,
and vanish for $H=0$, when $\Db(\rb,\rb') = \Db(\rb',\rb)$.
The effect of magnetic field is contained in the right-hand side (RHS) of
the boundary condition (\ref{eq:bcDrenorm}).
Since the difference $\Db_\nea -\Db_\sea$ is linear in $H$ for $\om_H\tau_0 \ll 1$,
it is linear in the factor $\om_H \tau_0 [1-c(\om)]$ in the RHS of Eq.~(\ref{eq:bcDrenorm}).
Combining this fact with Eq.~(\ref{eq:Dest}), we obtain

\beq
    \Db_\nea -\Db_\sea \propto \frac{1}{a^3} \frac{\om_H \tau_0 [1-c(\om)]}{D_0 [1-c(\om)]/a^2}
= \frac{1}{a^3} \frac{\om_H \tau_0 }{D_0/a^2},
\eeq
where the proportionality coefficient depends on the grain geometry only.
We see that the factors $[1-c(\om)]$ in the numerator and denominator
arising from the boundary condition (\ref{eq:bcDrenorm})
and differential equation (\ref{eq:Drenorm}), respectively,
cancel each other. Therefore, the Hall resistance $R_H$ [Eq.~(\ref{eq:RH})]
of the grain remains unaffected by weak localization  effects
and the correction $\de R_{H}^{WL}$ to it vanishes [Eq.~(\ref{eq:dRH})].
Consequently, the corresponding contributions to the Hall conductivity
and resistivity vanish:
\beq
    \frac{  \de \rho_{xy}^{(2)}(\om)}{ \rho_{xy}^{(0)}}
    =\frac{  \de \sig_{xy}^{(2)}(\om)}{ \sig_{xy}^{(0)}}=
    \frac{\de R_H^{WL}}{R_H} \equiv 0.
\label{eq:drhoxy2}
\eeq

\section{Results and conclusion}

Combining Eqs.~(\ref{eq:drhoxy1}) and
(\ref{eq:drhoxy2}),
we obtain that the first-order in the inverse tunneling conductance $1/g_T$
weak localization correction
$\de \rho_{xy}^{WL}=\de \rho_{xy}^{(1)}+\de \rho_{xy}^{(2)}$
to the Hall resistivity of a granular metal vanishes identically:
\beq
    \de \rho_{xy}^{WL}=0.
\label{eq:drhoxyWL}
\eeq

The weak localization correction
$\de \sig_{xy}^{WL}= \de \sig_{xy}^{(1)}+\de \sig_{xy}^{(2)}=\de \sig_{xy}^{(1)}$
[Eqs.~(\ref{eq:dsigxy1}), (\ref{eq:dsigxx}), and (\ref{eq:drhoxy2})]
to the Hall conductivity originates
from the renormalization of the tunneling conductance $G_T$ only,
the corresponding relative correction being
twice as large as that  to the longitudinal conductivity:
\[
    \frac{\de \sig_{xy}^{WL}}{\sig_{xy}^{(0)}}=
    2\frac{\de \sig_{xx}^{WL}}{\sig_{xx}^{(0)}}.
\]
The WL correction $\de \sig_{xx}^{WL}$
was studied in Refs.~\onlinecite{BelUD,BCTV,BVG}.


Whether the exact cancellation (\ref{eq:drhoxyWL}) obtained in the first order in $1/g_T$
is violated in higher 
orders or not remains a question of a separate
investigation\footnote{ To the best of our knowledge, we are
unaware whether the same question has been investigated for
homogeneously disordered metals.}. What is important, however, is
that in the same first order in $1/g_T$ (i) logarithmic
temperature-dependent corrections to both the longitudinal
$\rho_{xx}$\cite{ET,BELV}  and Hall
$\rho_{xy}$\cite{KEshort,KElong} resistivities due to Coulomb
interactions exist; (ii)  weak localization correction
$\de\rho_{xx}^{WL}(H)$ to $\rho_{xx}$ exists\cite{BelUD,BCTV,BVG},
being sensitive to the magnetic field\cite{BCTV,BVG}.
Therefore, we come to the conclusion that
in the leading order in $1/g_T$, in which quantum effects do come into play,
the effect of weak localization on the Hall resistivity is absent [Eq.~(\ref{eq:drhoxyWL})].

Experimentally,  our result (\ref{eq:drhoxyWL}) may be tested by  measuring
the dependence of the Hall coefficient $\rho_{xy}(H)/H$ on magnetic field $H$.
Since the weak localization correction $\de \rho_{xx}^{WL}(H)$
is sensitive to the magnetic field, Eq.~(\ref{eq:drhoxyWL}) states
that in the range of sufficiently low magnetic fields $H$, in which
the relative change in the longitudinal resistivity $\rho_{xx}(H)$ of the order of $1/g_T$
due to localization effects is predicted\cite{BCTV,BVG},
no comparable change in $\rho_{xy}(H)/H$ is expected.


In conclusion, we have studied the effects of weak localization on
Hall transport in granular metals. Calculating the first-order in
the inverse intergrain conductance corrections, we found that the
Hall resistivity of the system remains unaffected by weak
localization effects. This result is in agreement with the one
obtained for ordinary disordered metals. It holds for arbitrary
relevant values of temperatures $T$ and magnetic fields $H$, both
in the universal ``homogeneous'' regime of very low  $T$ and $H$
and in the ``structure-dependent'' regime of higher $T$ or $H$.

We thank Igor S. Beloborodov, Yuli V. Nazarov, and Anatoly F. Volkov for illuminating discussions
and acknowledge  financial support of Degussa AG (Germany),
SFB Transregio 12, the state of North-Rhine Westfalia, and the European Union.

\end{document}